\documentclass[journal=jacsat,manuscript=article]{achemso}

\usepackage{chemformula} 
\usepackage[T1]{fontenc} 
\usepackage{siunitx}
\usepackage{hyperref}

\author{Donato Conteduca}
\affiliation{School of Physics, Engineering and Technology, University of York, Heslington, YO10 5DD, York, UK}
\altaffiliation{These authors contributed equally to this work.}
\author{Saba N. Khan}
\affiliation{SUPA, School of Physics and Astronomy, University of St Andrews, North Haugh, St Andrews KY16 9SS, UK} 
\altaffiliation{These authors contributed equally to this work.}

\author{Manuel A. Mart\'inez Ruiz} 
\affiliation{SUPA, School of Physics and Astronomy, University of St Andrews, North Haugh, St Andrews KY16 9SS, UK} 
\author{Graham D. Bruce} 
\affiliation{SUPA, School of Physics and Astronomy, University of St Andrews, North Haugh, St Andrews KY16 9SS, UK} 
\author{Thomas F. Krauss} \email{ thomas.krauss@york.ac.uk}
\affiliation{School of Physics, Engineering and Technology, University of York, Heslington, YO10 5DD, York, UK} 

\author{Kishan Dholakia} \email{kishan.dholakia@adelaide.edu.au}
\affiliation{SUPA, School of Physics and Astronomy, University of St Andrews, North Haugh, St Andrews KY16 9SS, UK}
\alsoaffiliation{School of Biological Sciences, University of Adelaide, Adelaide, South Australia, 5005, Australia}
\alsoaffiliation{Centre of Light for Life, University of Adelaide, 5005, Australia}

\title[ ]
{Fano resonance-assisted all-dielectric array for enhanced near-field optical trapping of nanoparticles}


\begin{document}

\begin{abstract}
Near-field optics can overcome the diffraction limit by creating strong optical gradients to enable the trapping of nanoparticles. However, it remains challenging to achieve efficient stable trapping without heating and thermal effects. Dielectric structures have been used to address this issue, but they usually offer weak trap stiffness.
In this work, we exploit the Fano resonance effect in an all-dielectric quadrupole nanostructure to realize a twenty-fold enhancement of trap stiffness, compared to the off-resonance case. This enables a high effective trap stiffness of $1.19$  fN/nm for 100 nm diameter polystyrene nanoparticles with 4.2 mW/$\mu$m$^{2}$ illumination. 
Furthermore, we demonstrate the capability of the structure to simultaneously trap two particles at distinct locations within the nanostructure array. 
\end{abstract}
\maketitle



\section{Introduction}
The controlled optical manipulation of micrometer- and nanometer-sized particles remains a subject of interest across all of the natural sciences \cite{campugan2020optical,dholakia2011shaping,kotsifaki2019plasmonic, jones2015optical,conteduca2017photonic}.
Optical trapping requires a sufficiently strong gradient force on the associated particle. For the most commonly used dielectric particles (which have low absorption and associated thermal effects) the gradient force depends on the particle's polarizability, which scales with its volume. 
As a result of this scaling with polarizability, smaller particles require increasingly high intensities for stable trapping. For nanoparticles, however, the required optical power levels can be impractically high \cite{jones2015optical, Spesyvtseva2016May}.

Rather than increasing \emph{power}, one can choose to increase \emph{intensity}. But in the far field, we are restricted by the normal diffraction limit of light. Reverting to the near field, we can overcome this limitation by creating large gradients in the optical field on sub-wavelength dimensions. This means that we can enhance our capability to trap nanometre-scale dielectric particles in a controllable fashion. This realization has spawned the area of near-field optical trapping, which has seen a suite of methods developed to allow the confinement of objects well below a micron in size. \textcolor{black}{
To generate high-gradient fields, near-field methodologies employ photonic nanojets \cite{li2016trapping} and nanostructures \cite{kotsifaki2019plasmonic,xu2018optical} such as nanoapertures \cite{al2015label,pang2012optical,kotnala2014quantification}, nanoantennas \cite{xu2018optical,xu2019all}, metasurfaces \cite{kotsifaki2020fano}, and photonic crystal cavities \cite{descharmes2013observation,serey2012dna}. Depending on the material composition of these nanostructures, near-field nanotweezers can be categorized into two primary types: plasmonic (metallic) and dielectric (high refractive index) nanostructure-assisted configurations.}

\textcolor{black}{
Plasmonic traps generally give very high stiffnesses (approx. 0.1-9 fN/nm for particle size of 10-30 nm at a power density of 1mW/$\mu$m$^2$) \cite{kotsifaki2020fano,kotnala2014quantification}. However, significant heating is often reported\cite{ploschner2010optical}, which is deleterious to trapping in general and especially problematic if the sample is damaged by high temperatures. For instance, in Ref. \cite{kotsifaki2019plasmonic}, the need to switch off the laser (for 15 minutes between consecutive experimental runs) to ensure complete dissipation of heat is stated. The heating in plasmonic tweezers is largely governed by laser absorption in the metal layer rather than nanostructure design \cite{jiang2019temperature}.  
This issue can be mitigated by meticulous thermal management, as suggested in Ref. \cite{wang2011trapping}, or by transitioning to all-dielectric structures \cite{xu2018optical}. Although the all-silicon nanoantenna largely avoids heating, the trap stiffness is only $\sim$0.04 fN/nm for a 100 nm nanoparticle. Here, we seek a new all-dielectric nanostructure that will significantly improve trap stiffness while avoiding detrimental heating.}

In near field traps, resonant effects can be used to enhance the optical forces, e.g. the use of cavities in evanescent traps \cite{reece2006near,kotsifaki2020fano,conteduca2021exploring}. One such effect is the Fano resonance, which originates from the constructive and destructive interference of a narrow discrete resonance with a broad spectral line or continuum. This results in an asymmetric line shape and has found applications in many areas, limited not only to optical trapping but also in Raman spectroscopy for molecular detection\cite{le2008metallic,huang2019sers,wu2012fano}, label-free biosensing \cite{yesilkoy2019ultrasensitive,conteduca2021dielectric}, sub-Doppler laser cooling of atoms \cite{Bruce2017Apr} and topological photonics \cite{limonov2017fano}. 

In this paper, we experimentally validate for the first time the use of the Fano resonance for trapping in an all-dielectric nanostructure.  Our work focuses on the confinement of 100 nm diameter nanoparticles and shows an enhancement in excess of twenty in the trap stiffness in comparison to that achieved by nonresonant dielectric Si-nanoantennas \cite{xu2018optical}. Exploiting the Fano resonance allows dynamic control of the optical forces and trapping efficiency by the polarization state of the excitation light beam \cite{zaman2022dynamically}. We also report observations of multiple trapping with two particles held at adjacent sites, a feature that is likely to find wide utility.

\section {Quadrupole array: design and fabrication}
\begin{figure}[h!]
 \centering
\includegraphics[width=0.9\linewidth]{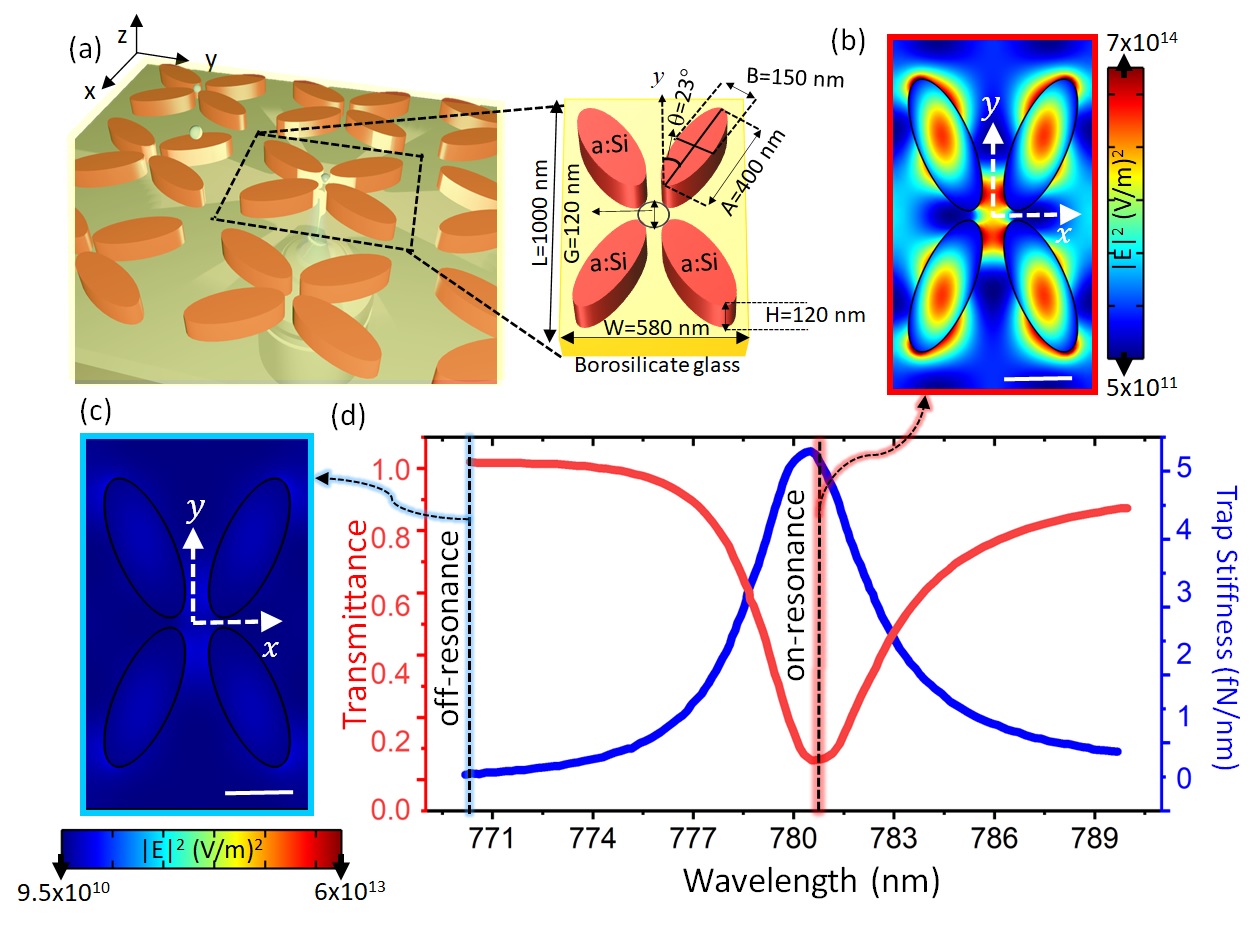} 
\caption{\textcolor{black}{
(a) An array of quadrupoles produces near-field confinement of laser light which can be used to trap nanoparticles. The cutout shows the geometrical parameters of the quadrupole unit cell. The electric field distribution in the x-y plane for (b) a resonantly excited nanostructure and (c) an off-resonantly excited nanostructure. (d) The transmission of light through the nanostructure and the trap stiffness for a 100nm polystyrene particle as functions of the excitation wavelength, demonstrating a Fano resonance shape. The scale bar shows 200 nm and is  common to panels (b) and (c).}}
\label{fig: simulation}
\end{figure}

The proposed dielectric nanostructure consists of an array of amorphous silicon quadrupoles $\text{(a-Si:H}$; having refractive index \mbox{$n$ = 2.4} and absorption coefficient $k$ $= 5\times 10^{-4}$ at wavelength $\lambda$ = 785 nm), embedded in low refractive index glass substrate ($n_{sub}$ = 1.45). The quadrupole unit cell is composed of four elliptical meta-atoms arranged in a mirrored pair along the vertical (or horizontal) direction  (see Fig.\ref{fig: simulation}a). The period along the $x$ and $y$ axes of the unit cell are W = 580 nm, and L = 1000 nm respectively. The geometrical parameters of the elliptical meta-atoms are: thickness H = 120 nm, the short and long axes are B = 150 nm 
and A = 400 nm 
respectively. The orientation of the meta-atom with respect to the $y$ axis is characterized by the rotation angle $\theta$, which is introduced to provide an asymmetry that allows polarization-sensitive localization of the electromagnetic field in the quadrupole array. The quadrupole meta-atoms are positioned to accommodate a central gap region (G $\sim120$ nm) slightly larger than the nanoparticles for which the trap is designed (diameter $\sim100$ nm). 

To increase the light confinement within the unit cell by exploiting Fano resonance, the nanostructure is illuminated by laser light propagating in the $z$ direction with a spatial extent sufficient to cover multiple unit cells
. A finite element package (FEM; COMSOL Multiphysics) was used to numerically calculate the optical responses of the quadrupole array, assuming an infinite array and linearly polarized (x-polarized) plane wave illumination. The quadrupole design is inspired by ref. \cite{liu2018extreme}. \textcolor{black}{The quadrupole structure supports a quasi-bound state in the continuum (quasi-BIC) mode for which the optical response, in terms of Q-factor and resonance amplitude, is dependent on the geometrical parameters of the array (see section S1 to S5 in supplementary information for more details on the design aspects of the quadrupole structure)}. 
The design harnesses the array configuration to exhibit strong near-field confinement while optimizing the Q-factor. Whilst a high Q-factor seems desirable as a larger energy enhancement will enable a stiffer trap, a very high Q-factor can be undesirable as the resonance detunes when the nanoparticle enters the trap. If this detuning is larger than the resonance linewidth, this would detune the trap off-resonance altogether\cite{conteduca2021exploring}. To balance these considerations, we have designed a quadrupole array that possesses a resonance linewidth (approx. FWHM = 5 nm) slightly larger than the spectral shift (approx. 1 to 2 nm) envisaged in the trapping event, yielding a Q-factor of 140. 
\begin{figure}[ht!]
 \centering
\includegraphics[width=0.85\linewidth]{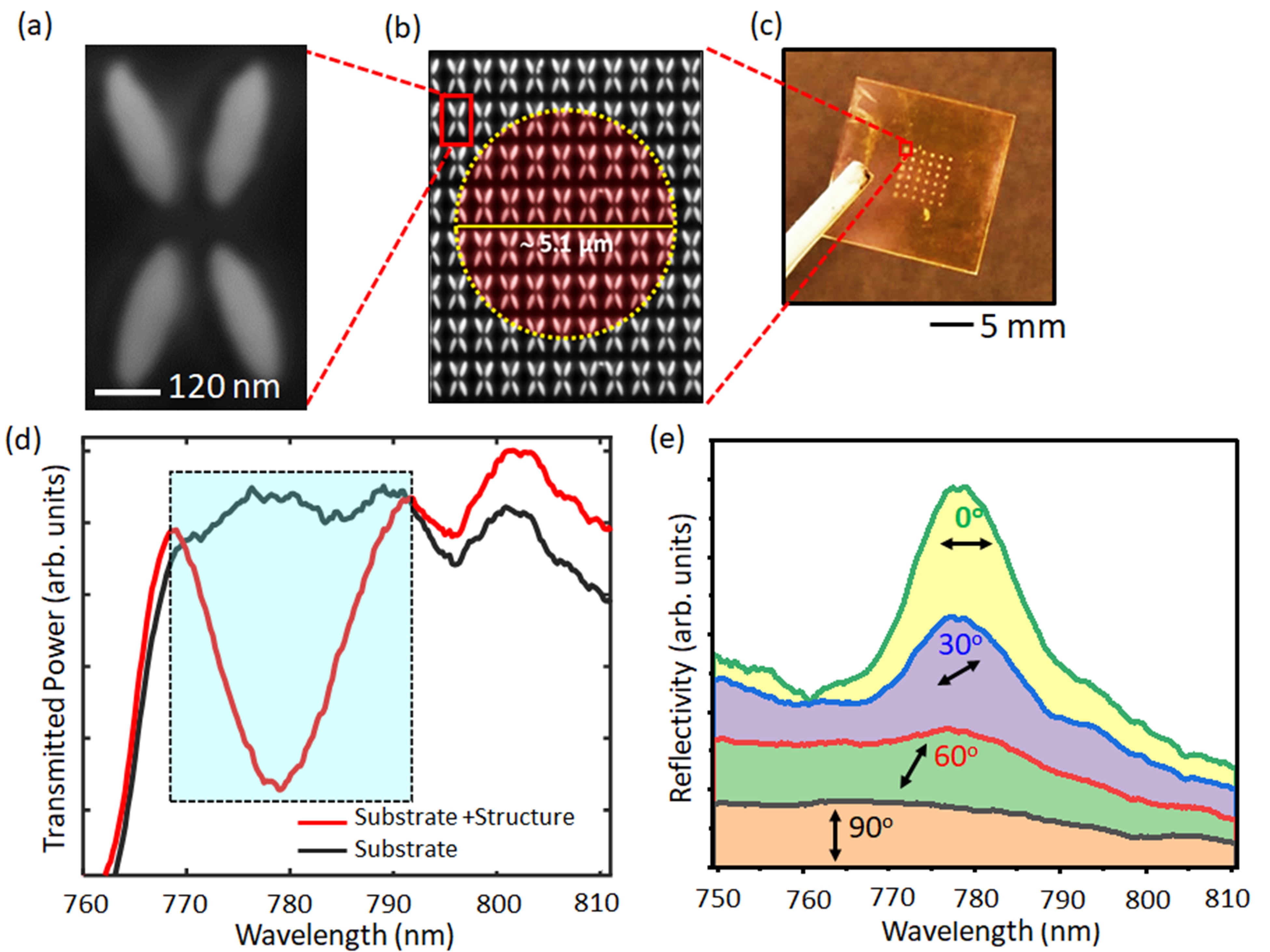}
\caption{(a) Scanning electron microscope (SEM) image of the quadrupole meta-unit and (b) array. (c) Optical micrograph of the metasurface structure on the dielectric substrate. (d) Transmission spectrum of the dielectric substrate (black) and the nanostructure (red), which shows increased light confinement at a resonance peak of 779.1 $\pm$ 0.2 nm. The resonance characteristic of the nanostructure was obtained by illuminating 6 $\times$ 5 quadrupole units in deionized water. (e) The input polarization dependence of the Fano resonance characteristic of the quadrupole nanostructures suggests that the trapping forces can be dynamically controlled by the input state-of-polarization of the excitation beam.} 
\label{fig2:Fabrication}
\end{figure}
\textcolor{black}{The electric field distribution in the cross-section of the quadrupole unit cell is shown in Fig.\ref{fig: simulation}(b). A particular feature of this design is that the light field is highly localized around the central gap region, which is desirable to avoid particles sticking to the nanostructure. As shown in  Fig.\ref{fig: simulation}(b), the design creates two localised maxima at the narrowest gaps between the meta-atoms. These maxima are separated by $\sim100$nm along the $y$-axis, so that a 100 nm-diameter particle is equally pulled between both maxima and held maximally separated from the meta-atoms. The design of the intensity profile is reminiscent of offset-focus dual beam optical traps \cite{thalhammer2011}, which are advantageous for trapping objects which may be thermally damaged in the highest intensity regions of the trapping field.}

\textcolor{black}{As shown in Fig.~\ref{fig: simulation}c, far from the resonance, the structure does not exhibit strong light confinement.} The transmission spectrum of the nanostructure is dominated by the Fano resonance effect, as shown in red in Fig.\ref{fig: simulation}(d). \textcolor{black}{The increased field confinement due to the resonance is illustrated in Fig.\ref{fig: simulation}(b) and (c), where excitation of the quadrupole with a wavelength detuned by 10 nm reduces the peak intensity by three orders of magnitude when compared to the resonant case. This can also be seen in the simulated optical trap stiffness for a 100 nm polystyrene nanoparticle with 5 mW/$\mu$m$^{2}$ illumination intensity (blue curve in  Fig.\ref{fig: simulation}(d)).} The Fano resonance characteristics of the quadrupole array can be spectrally tuned to a desired wavelength by judicious choice of the geometric parameters of the quadrupole array, while the Fano resonance efficiency is affected by the size of the array and the beam's state of polarization. These dependencies on geometrical parameters and the state of polarization of the excitation beams are detailed in sections S1 and S2 of the supplementary information. \textcolor{black}{The field confinement is dramatically altered by the resonance created by the repeated structure. In section S4 of the supplementary information, we show the field confinement realised by a decoupled single meta-atom, which produces a peak field amplitude weaker by a factor of 10, and also with a markedly different spatial distribution which is less suited to trapping in the center of the quadrupole structure.}

The nanostructure is fabricated using e-beam lithography, by nanostructuring a 120 nm thick amorphous Si film on a glass substrate, followed by dry etching (see section S6 in supplementary information for the details of the fabrication of the quadrupole nanostructures). Scanning Electron Microscope (SEM) micrographs of a quadrupole unit cell and the array are shown in Fig.\ref{fig2:Fabrication} (a) and (b) respectively, and the final nanostructure and substrate are shown in Fig. \ref{fig2:Fabrication}(c). 

To characterize the resonance behavior of the nanostructure, the structure was immersed in deionized water and an array of 6 x 5 meta-units was illuminated with white light (OSL 1 Fiber Illuminator, Thorlabs). The transmission spectrum is recorded in the presence and absence of the nanostructure to extract the Fano resonance characteristics. Figure \ref{fig2:Fabrication}(d)  shows the signature of the Fano resonance: a pronounced reduction in transmission which is centered at 779.1 $\pm$ 0.2 nm with a Q-factor of $\sim65$. \textcolor{black}{The resonance wavelength can be tuned by a few nm by illuminating different areas of the structure, as detailed in supplementary section S7.} Figure \ref{fig2:Fabrication}(e) shows the reflectance curve of the quadrupole array as a function of the polarization angle. For vertical polarization (represented by the black reflectivity curve in Fig. \ref{fig2:Fabrication}(d)), the quadrupole nanostructure lacks Fano resonant spectral features. The efficiency of light confinement was seen to increase with changing polarization angles, showing maximum light-field confinement when the excitation light is horizontally polarized. A similar trend is observed in the numerical simulation (see Fig. S2 in the supplementary information). This indicates that the mode of the quadrupole array is formed by the collective interaction between the individual meta-atoms, and the polarization of the excitation beam can serve as a tool to dynamically control the trap stiffness of the nanoparticles. Further verification of the critical contribution of the array to the light confinement was performed by measuring the reflectance spectrum of a decoupled structure with an identical meta-unit but a much larger periodicity, which does not show any resonance effect (see section S8 of the supplementary document). 

\section{Near-field optical trapping}

\begin{figure}[ht!]
 \centering
\includegraphics[width=0.85\linewidth]{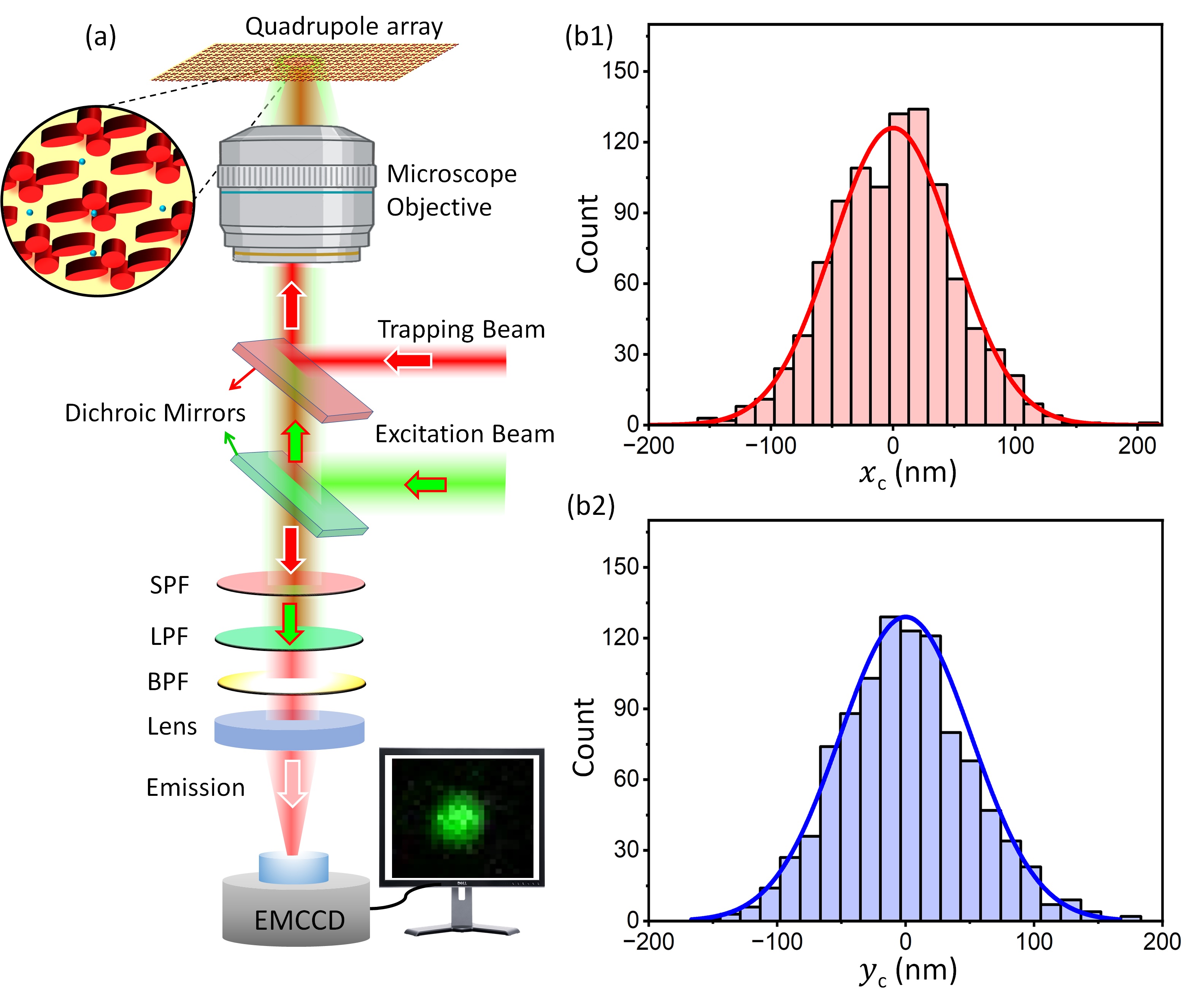}
\caption{(a) Simplified schematic of the experimental setup, abbreviations are- BPF: band pass filter, LPF: long pass filter, SPF: short pass filter; EM-CCD: electron-multiplying charged coupled device. (b1 and b2) Brownian motion histogram of a trapped particle. Position histograms computed from 1000 EM-CCD frames of a trapped 100 nm nanoparticle, relative to the trap center along x- and y-axes. The illumination power density was I = 5.1 mW/$\mu m^2$.  } 
\label{fig3:expsetup}
\end{figure}

To use the quadrupole array for particle trapping, it was illuminated by a tunable wavelength, continuous-wave Ti: sapphire laser (Coherent MIRA 900-F). The laser beam was weakly focused using a low-numerical aperture microscope objective lens (NA = 0.3, 40X, Nikon) to excite 6x5 quadrupole units of the array (see Fig. \ref{fig3:expsetup}(a)). 
\textcolor{black}{The presence of a trapped particle shifts the nanostructure resonance and thereby changes the intensity of }the transmitted laser light\textcolor{black}{. This transmitted light was therefore} collected through a long working distance microscope objective (50X, Mitutoyo) and \textcolor{black}{monitored using }an avalanche photodiode (APD410A/M, Thorlabs). The supplementary document provides a comprehensive description of the experimental setup employed to achieve Fano resonance assisted near-field trapping using the quadrupole nanostructures (see section S9 of the supplementary information).  
The trapping process was also investigated by performing fluorescence microscopy on red fluorescent nanoparticles (R100, Duke Scientific Corp.). To do so, a weakly focused nanosecond laser (SPOT 10-200-532, Elforlight) operating at 532 nm with 1 $\mu$J energy and repetition rate of 20 kHz was introduced in the same microscope arrangement over a wide field of view for fluorescence excitation. The emitted fluorescent signal from the red fluorescent polystyrene nanoparticles was detected using an electron-multiplying charged coupled device (EMCCD; iXon Ultra 897, Andor, Oxford Instruments) after passing through a narrow band pass emission filter (FF01-640/40-25, Semrock).

Dielectric polystyrene nanoparticles of 100 $\pm$ 6 nm diameter were diluted in heavy water with a volume concentration of 0.05 $\%$. A small amount of Tween-20 surfactant was added with a volume concentration of 0.1$\%$ to the particle solution, and the final solution was sonicated to prevent the formation of aggregates. The chamber preparation for realizing near-field trapping with a quadrupole array is further detailed in section S9 of the supplementary information. The effective trap stiffnesses along the two orthogonal directions are obtained from the variances in the position of the trapped nanoparticle using $\kappa_{i} = \frac{\text{k}_{B}\text{T}}{\text{var(i)}}$ for $i = x$ or $y$ \cite{tanaka2013nanostructured}. Here, 1$\text{k}_B$$\text{T}$ = 4.05 × $10^{-21}$ J. 
Notably, the potential well produced by the quadrupole array is not strictly harmonic {(see Fig. S5 of the supplementary information)}, and therefore an effective trap stiffness was measured considering an ideal harmonic trap that produces the localization (position variance) similar to our all-dielectric quadrupole array. Panels (b1) and (b2) in Fig. \ref{fig3:expsetup} present typical histogram results depicting the Brownian movement of the centroid of a nanoparticle at a power density of 5.1 mW/$\mu m^{2}$. Each data point was obtained by extracting the centroid of the fluorescent emission extracted from 1000 frames. The trap stiffnesses along the $\it{x}$ and $\it{y}$ directions are calculated to be $1.63\pm 0.22$  fN/nm and $1.54 \pm 0.25$  fN/nm respectively. We were able to maintain the particle in the trap for hours without requiring any cycling. The simulation predicts that the heating in the quadrupole array is two orders lower in magnitude than the plasmonic structures which permits higher intensity to be used for stable traps (see section S3 of the supplementary information).

\begin{figure}[ht!]
 \centering
\includegraphics[width=0.9\linewidth]{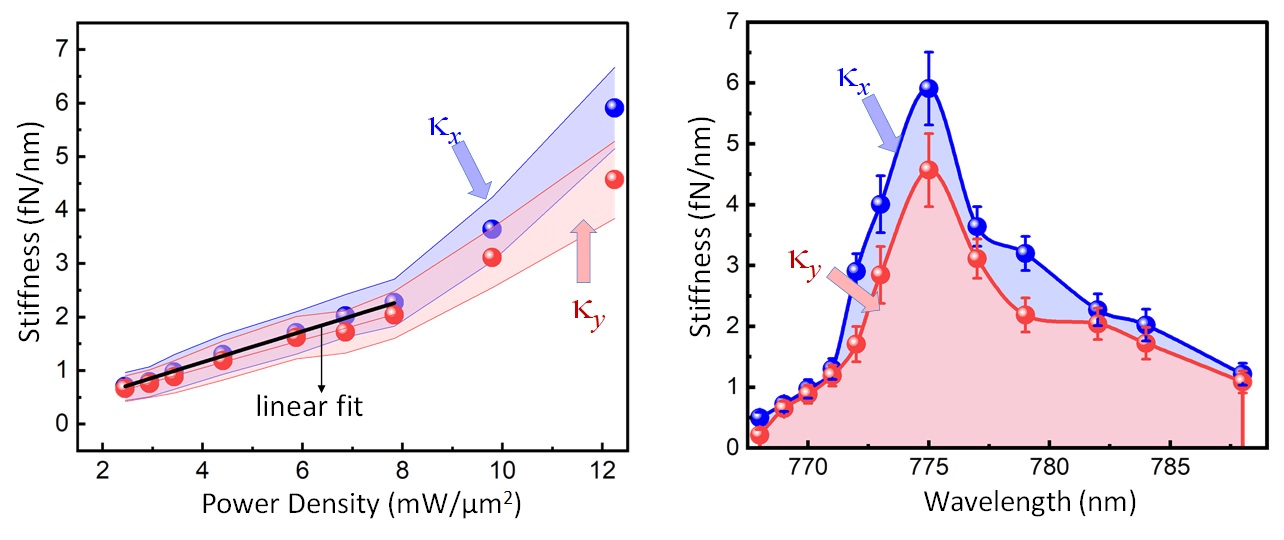}
\caption{Trap-stiffness of a single 100 nm polystyrene nanoparticle as a function of trapping laser (a) power density, and (b) wavelength. Experimentally obtained stiffness values are computed from Brownian motion histograms such as those in Fig. \ref{fig3:expsetup}(b1 and b2). Error bars denote the standard deviation in the trap stiffness measurements over 3 repetitions. By appropriate selection of the laser wavelength, the trap stiffness can be enhanced by a factor of 20.
} 
\label{fig4:wavelength_and_power}
\end{figure}
It is obvious that increasing the laser intensity should increase the effective trap stiffness, but the behavior was not strictly linear as would have been expected in conventional tweezers (see Fig. \ref{fig4:wavelength_and_power}(a)). This could be due to the following reasons: (i) the potential arising from the field localization was not strictly harmonic, (ii) the trapped nanoparticle cannot move freely in all directions due to the closely packed geometry of the quadrupole unit cell, (iii) the presence of particle-surface interactions (iv) convection and thermophoresis effects at higher intensities. Note that all these measurements are corresponding to a horizontally polarized trapping beam which provides the maximal energy enhancement with the quadrupole nanostructure. 
\textcolor{black}{The trap stiffness follows the expected Fano-shaped behavior as a function of wavelength, as shown in }
Fig.\ref{fig4:wavelength_and_power}(b) 
\textcolor{black}{where the laser intensity is 12 mW/$\mu m^{2}$. The peak trap stiffness is a factor of 20 higher the trap stiffness when the laser is tuned far from the Fano resonance, and approximately 25 times higher than that of an all-dielectric silicon nanoantenna \cite{xu2018optical}. Movie S1 in the supplementary information demonstrates the nanoparticle arrival at the near-field trap. As the video progresses, you can observe a reduction in trap stiffness with changing polarization.} 
\begin{figure}[ht!]
 \centering
\includegraphics[width=0.9\linewidth]{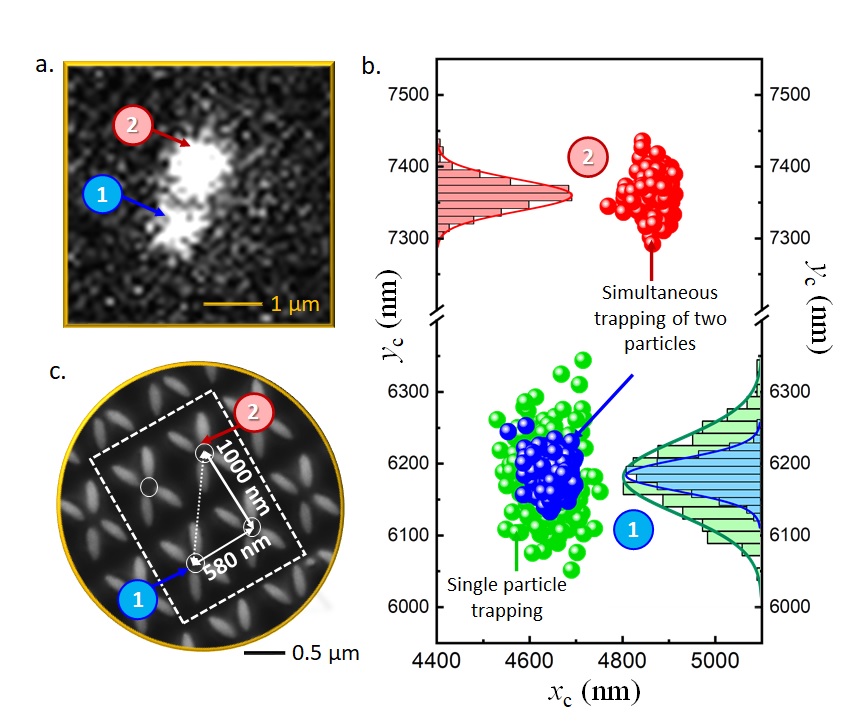}
\caption{Simultaneous trapping of two particles at the adjacent trapping sites. (a) camera frame showing the fluorescent signal of the individual particles. (b) Scatter plot showing Brownian motion of the particles (with respective histograms) in the trap. (c) SEM image showing the adjacent site separation and the probable trapping sites (labeled as 1 and 2) within the nanostructure array. } 
\label{fig5:simulatneoustrapping}
\end{figure}
{\\}
In Fig. \ref{fig5:simulatneoustrapping} and Supplementary Movie S2, we show an example of the simultaneous trapping of two individual, 100 nm diameter, red-fluorescent nanoparticles at adjacent (diagonal) trapping sites, with an illumination intensity of 7.8 mW/$\mu m^{2}$. 
The Brownian motion of the particles was analyzed using the EMCCD-frames corresponding to individual particles. 
\textcolor{black}{In the example shown here, initially a single particle was trapped at site 1 (blue points in Fig. \ref{fig5:simulatneoustrapping}(b). From measurements of the position of the particle taken over 1000 frames, we extract a mean trap stiffness of 1.9 fN/nm. At the end of this period, a second particle became trapped at site 2 (red points in Fig. \ref{fig5:simulatneoustrapping}(b)). The centres of the two particles' position histograms give a separation of $1184\pm 16$ nm, which is consistent with the $1156$ nm separation expected for two diagonally offset trapping sites. The trap stiffness at this second site was determined to be 6.9 fN/nm. Importantly, the trap stiffness of the first particle at site 1 instantaneously increased due to the presence of the second particle, as can be seen in the blue points in Fig. \ref{fig5:simulatneoustrapping}(b). The trap stiffness at site 1 when a second particle was trapped at site 2 was 7.7 fN/nm. This increase in trap stiffness is indicative of dielectric loading effects commonly seen in self-induced back-action trapping \cite{juan09}, whereby the presence of a particle in the trap shifts the resonance wavelength and modifies the light confinement across the whole array. This opens prospects not only for multi-particle trapping, but also for studies of collective dyanmics and synchronisation.}

\section{Conclusions}\label{5}
In this work, we have demonstrated near-field trapping using a Fano resonance-assisted all-dielectric quadrupole nanostructure. We observe that trap stiffness was enhanced by approximately twenty-fold when the trapping laser was tuned on resonance compared to an off-resonance wavelength. The efficiency of the Fano resonance effect can be dynamically controlled by the polarisation of the excitation beam which can be used as a tool to control trap stiffness. The trap stiffness achieved with the quadrupole array was an order (25-fold) higher than the non-resonant Si nanoantennas \cite{xu2018optical} not exhibiting the Fano resonance effect. Such a system may 
be of relevance for Raman spectroscopic or other analyses of trapped nanometric particles (e.g. viruses) \cite{ashok2012optical}. The array structure also enables simultaneous trapping of multiple individual particles. We presented an initial outcome demonstrating the trapping of two particles within the quadrupole nanostructure. Additionally, we illustrated how the presence of a second particle alters the trapping potential experienced by the first particle.  
These interactions of two or more particles in the array will be a topic of future study, as they may open new frontiers in optical binding, synchronization, and sympathetic control over particle motion \cite{kotar2013optimal,arita2022all}.  
\section{Supporting Information}
The supplementary document offers a thorough description of the simulation performed to analyze the optical response of the quadrupole array, the fabrication process of the nanostructure device, and the experimental setup utilized to achieve near-field trapping through the utilization of Fano resonance effect. 
\section*{Funding sources}
This work was supported by the UK Engineering and Physical Sciences Research Council (EP/P030017/1), and the Australian Research Council (grant DP220102303).  
\section{Acknowledgements}
We thank Paloma Rodr\'iguez Sevilla for her early contributions to the experimental setup and methods and Yoshihiko Arita for useful discussions.
\bibliography{sample}
\end{document}